\documentclass{aastex62}
\usepackage{graphicx}
\begin{document}

\title{High spectral resolution observations toward Orion BN at 6~$\mu$m: no evidence for hot water.}

\author{Nick Indriolo}
\affil{Space Telescope Science Institute, Baltimore, MD 21218, USA}
\author{Jonathan C. Tan}
\affil{Department of Space, Earth \& Environment, Chalmers University of Technology, SE-412 93 Gothenburg, Sweden}
\affil{Department of Astronomy, University of Virginia, Charlottesville, VA, 22904, USA}
\author{A. C. A. Boogert}
\affil{Institute for Astronomy, University of Hawaii at Manoa, Honolulu, HI, 96822, USA}
\author{C. N. DeWitt}
\affil{USRA, SOFIA, NASA Ames Research Center MS 232-11, Moffett Field, CA 94035, USA}
\author{E. J. Montiel}
\affil{Department of Physics, University of California Davis, Davis, CA, 95616, USA}
\author{D. A. Neufeld}
\affil{Department of Physics \& Astronomy, Johns Hopkins University, Baltimore, MD 21218, USA}
\author{M. J. Richter}
\affil{Department of Physics, University of California Davis, Davis, CA, 95616, USA}

\begin{abstract}
Orion BN has a large proper motion and radial velocity with respect to the gas and other stars in the region where it is presumed to have formed. Multiple dynamical interaction scenarios have been proposed to explain this motion. In one case BN is thought to have interacted with stars in the Trapezium cluster, while in another it is thought to have interacted with source I while deeply embedded in molecular gas. If there is dense gas that has been retained in close proximity to BN, it may be evidence that the latter scenario is favored. We observed BN at high spectral resolution in three windows near 6~$\mu$m using SOFIA/EXES targeting the $\nu_2$ vibrational band of H$_2$O. Absorption from only three transitions of H$_2$O is detected, and through kinematic analysis is associated with cool, dense foreground gas, not BN itself. We find no evidence for H$_2$O absorption or emission at the systemic velocity of BN.
\end{abstract}

\section{Introduction} \label{sec_intro}

The Becklin-Neugebauer (BN) object \citep{becklin1967} is a young, massive \citep[8.0--$12.6$~$M_{\odot}$;][]{scoville1983,rodriguez2005} star, with fast motion through the Orion Nebular Cluster (ONC). It has a radial velocity \citep[$v_{\rm LSR}=23.5\pm0.5$~km~s$^{-1}$;][]{plambeck2013} that differs from that of the Orion molecular cloud \citep[$v_{\rm LSR}\sim9$~km~s$^{-1}$;][]{melnick2010}, and a high proper motion \citep[$12.6\pm0.6$~mas~yr$^{-1}$; ][]{rodriguez2005}. This amounts to BN having a 3D velocity of about 30~km~s$^{-1}$ relative to its surroundings \citep[assuming a distance of $\sim400$~pc;][]{menten2007,kounkel2017}, the origin of which is a matter of debate. Two scenarios have been proposed to explain the motion of BN. The first is dynamical ejection from the $\theta^1{\rm Ori\:C}$ system (now a binary) in the Trapezium region near the center of the ONC about 4,000 years ago \citep{tan2004}. Current properties of the $\theta^1{\rm Ori\:C}$ system such as orbital binding energy and recoil proper motion can be explained if BN was ejected from the system \citep[see $N$-body simulations of][]{chatterjee2012}. Alternatively,  \citet{bally2005} and \citet{rodriguez2005} proposed that dynamical interaction of BN, radio source I, and a potential third body (originally proposed to be radio source n) could have produced the high proper motions of BN and source I in roughly opposite directions. \citet{luhman2017} have found that the third star in this scenario is very likely to be source x, based on an HST-detected proper motion of this source. \citet{farias2018} carried out $N$-body modeling of an ejection of BN and source x from source I, which would then be a hardened or merged binary, concluding that source I would need to be $\sim20\:{M}_\odot$, consistent with recent ALMA observations \citep{bally2017,ginsburg2018}.


The ejection or runaway nature of BN is likely connected to the apparently ``explosive'' outflow from the KL nebula \citep{allen1993}. However, there are a number of uncertainties still associated with the above scenarios. In particular, how exactly the ``explosive'' outflow is launched during the interaction and whether all of the stars---source I, BN, and source x---were originally forming from the KL nebula (i.e., the Orion Hot Core). The presence of dense gas around BN can help constrain these scenarios. For example, if BN was in the process of forming from the KL nebula, then dense gas of its inner accretion disk should have been retained by BN even after ejection. Here we present high spectral resolution observations of BN made at 6~$\mu$m covering the $\nu_2$ ro-vibrational band of H$_2$O, which has previously been used to identify hot, dense gas in close proximity to massive protostars \citep[e.g.,][]{boonman2003,indriolo2015exes}.

\section{Observations and Data Reduction}

Orion-BN was observed using the Echelon-Cross-Echelle Spectrograph \citep[EXES;][]{richter2010} on board the Stratospheric Observatory for Infrared Astronomy \citep[SOFIA;][]{young2012} on 2017 January 26 (UT).  Spectra were acquired in cross-dispersed high-resolution mode targeting central wavenumbers of 1485.24~cm$^{-1}$, 1639.29~cm$^{-1}$, and 1747.25~cm$^{-1}$ (hereafter referred to as the 6.7~$\mu$m, 6.1~$\mu$m, and 5.7~$\mu$m spectra, respectively). The 6.7~$\mu$m spectrum was obtained at an altitude of 37,930~ft (11,561~m), while the 6.1~$\mu$m and 5.7~$\mu$m spectra were both obtained at an altitude of 40,000~ft (13,106~m). The entrance slit had a width of 1\farcs24, providing a resolving power (resolution) of $\sim85$,000 (3.5~km~s$^{-1}$), and a length of about 10\arcsec\ (varies slightly between settings). To facilitate the removal of telluric emission lines, exposures alternated between on-target and a blank sky position 43\arcsec\ away. Total, on-target exposure times for the 6.7~$\mu$m, 6.1~$\mu$m, and 5.7~$\mu$m spectra were 660~s, 798~s, and 300~s, respectively. Sirius was observed at the same three spectral settings and using the same strategy for use as a telluric standard star.
 
\begin{figure}[h]
\epsscale{1.2}
\plotone{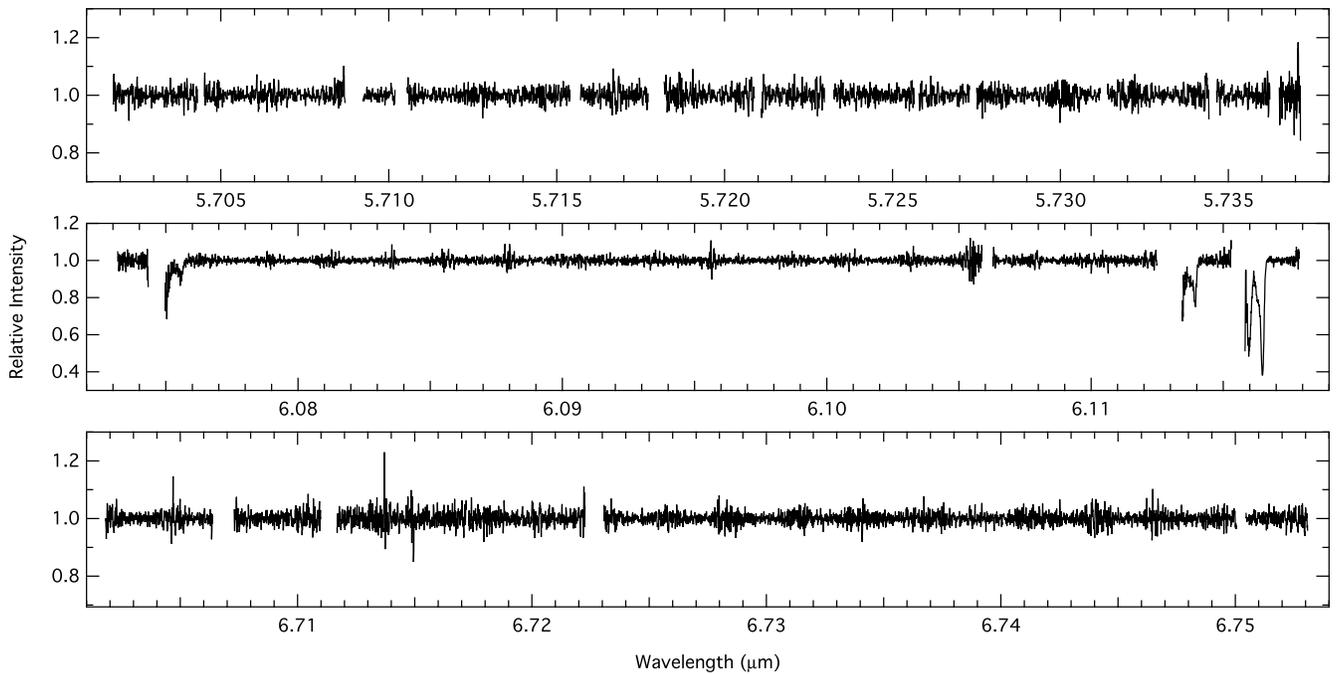}
\caption{Spectra of Orion~BN over the full wavelength range covered by our SOFIA/EXES observations. Each spectrum corresponds to a separate observation at a different grating setting. Oscillations in the noise level result from the cross-dispersed nature of the instrument, with higher noise regions coming from the edges of the individual echelle orders. Gaps in the spectra correspond to regions where telluric water vapor lines reduce transmission levels to near zero. The water absorption lines discussed herein are all from the spectrum in the middle panel.}
\label{fig_spectra}
\end{figure}

Data were processed using the {\tt Redux} pipeline \citep{clarke2014} with the {\tt fspextool} software package---a modification of the Spextool package \citep{cushing2004}---which performs source profile construction, extraction and background aperture definition, optimal extraction, and wavelength calibration for EXES data.
We used this software to produce wavelength calibrated spectra for each individual order of the echellogram. These individual spectra were then stitched together using an average of both orders in the overlap regions to produce a continuous spectrum for each of the three separate observations. To remove baseline fluctuations and atmospheric features the spectra of BN were divided by the corresponding spectra of Sirius using custom macros developed in IGOR Pro\footnote{https://www.wavemetrics.com} that allow for interactive scaling of the atmospheric features in the standard star spectrum to best match those in the science target \citep{mccall2001}. The resulting ratioed spectra were then divided by a 30 pixel boxcar average of the continuum level (extrapolated across absorption lines) to produce the normalized spectra shown in Figure \ref{fig_spectra}. 


\section{Analysis}

\begin{figure}
\epsscale{0.65}
\plotone{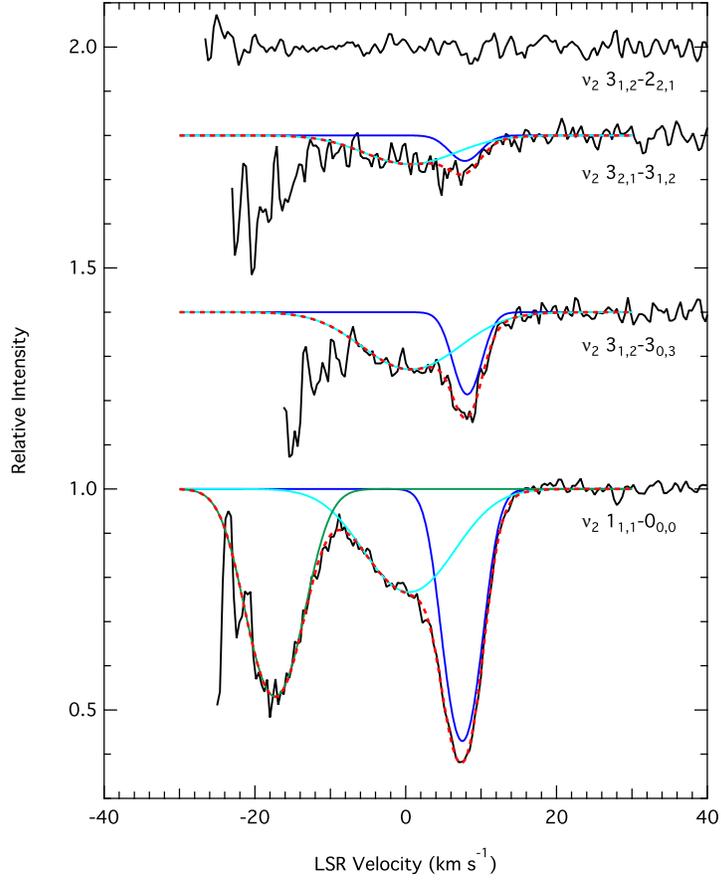}
\caption{Spectra showing the water absorption lines observed toward Orion~BN. Fits to the components at 8~km~s$^{-1}$, 0.5~km~s$^{-1}$, and $-$17~km~s$^{-1}$, are shown by blue, cyan, and green curves, respectively, and the sum of all components is shown as a red, dashed curve. The top spectrum shows a non-detection of the transition predicted to have the strongest absorption after the detected transitions in the observed wavelength range ($\nu_2$ 3$_{1,2}$--2$_{2,1}$ at 6.1068262~$\mu$m). Spectra have been shifted vertically for clarity.}
\label{fig_spectra_fits}
\end{figure}

Only the 6.1~$\mu$m spectrum (middle panel of Figure \ref{fig_spectra}) shows absorption features toward BN; the other two spectra are featureless. Signal-to-noise (S/N) levels on the continuum vary from about 100 to 40 depending on whether the spectrum was observed near the center or edge of an echelle order, respectively. The three detected absorption features are presented in Figure \ref{fig_spectra_fits}, shown in the LSR velocity frame for their respective transition wavelengths, and all show multiple absorption components. Absorption from the $\nu_2$ 1$_{1,1}$--0$_{0,0}$ water transition out of the ground rotational state is the strongest, with components at 8~km~s$^{-1}$, 0.5~km~s$^{-1}$, and $-17$~km~s$^{-1}$. The other two transitions also show absorption components at 8~km~s$^{-1}$ and 0.5~km~s$^{-1}$, but the $-17$~km~s$^{-1}$ component in those transitions fell in regions of low atmospheric transmission or poor atmospheric removal. The transmission spectra in Figure \ref{fig_spectra_fits} were fit using a sum of gaussian functions defined in terms of optical depth. Each individual fit component is shown by a solid color curve, and the overall fit to the absorption profile is shown as a red dashed curve. The resulting fit parameters are presented in Table \ref{tbl_measurements}. Note that for the weak absorption of the $\nu_2$ 3$_{2,1}$--3$_{1,2}$ transition the line center velocity and gaussian full width at half maximum (FWHM) for the two absorption components were forced to be the average of those parameters measured for each component from the other two transitions.

\begin{figure}
\epsscale{0.6}
\plotone{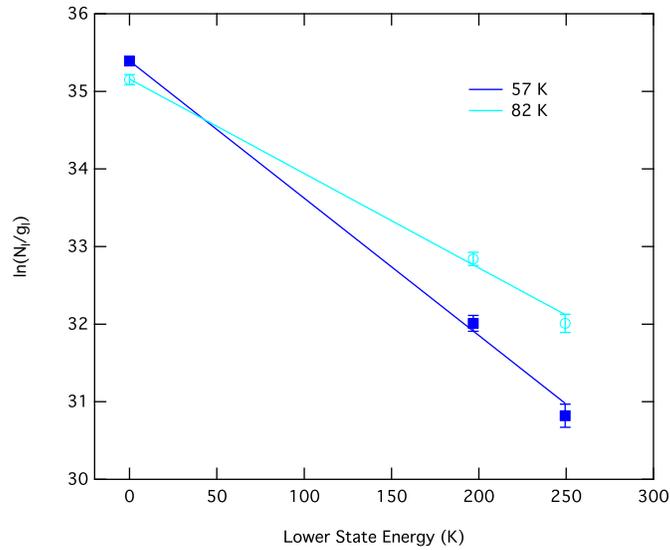}
\caption{Rotation diagrams for the 8~km~s$^{-1}$ component (blue) and 0.5~km~s$^{-1}$ component (cyan). Color coding corresponds to the gaussian fits shown in Figure \ref{fig_spectra_fits}.}
\label{fig_rotdiag}
\end{figure}

Integrated optical depths ($\int\tau dv$) for each component were computed as the area of the gaussian fitting functions, and are also presented in Table \ref{tbl_measurements}. Assuming optically thin absorption (reasonable given $\tau_0 < 1$ and the 3.5 km~s$^{-1}$ velocity resolution) column densities in the lower states of the observed transitions are computed as
\begin{equation}
N_l= \frac{g_{l}}{g_{u}}\frac{8\pi}{A_{ul}\lambda^3} \int\tau dv,
\label{eq_column}
\end{equation}
where $g_l$ and $g_u$ are the lower and upper state statistical weights, $A_{ul}$ is the transition spontaneous emission coefficient, and $\lambda$ is the transition wavelength. Using the state specific column densities we construct rotation diagrams for the two velocity components where three transitions are observed (8~km~s$^{-1}$ and 0.5~km~s$^{-1}$). These are presented in Figure  \ref{fig_rotdiag}. Assuming local thermodynamic equilibrium, the 8~km~s$^{-1}$ component has a rotational temperature of $T=57\pm 2$~K, and a total water column density of $N({\rm H_{2}O})=(3.67\pm0.22)\times10^{16}$~cm$^{-2}$, while the 0.5~km~s$^{-1}$ component has $T=82\pm 3$~K and $N({\rm H_{2}O})=(4.91\pm0.39)\times10^{16}$~cm$^{-2}$.

\section{Discussion}

\subsection{Comparison to Previous Near-to-Mid Infrared Observations of BN}

Orion~BN has been the target of several observing campaigns. All 3 of the velocity components that we observe in H$_2$O absorption have previously been identified\footnote{We assume here that the $-3$~km~s$^{-1}$ component reported in CO absorption corresponds to the 0.5~km~s$^{-1}$ component identified herein.} in absorption lines from the $v=1$--0 bands of $^{12}$CO and $^{13}$CO at 4.7~$\mu$m and the $v=2$--0 band of $^{12}$CO at 2.3~$\mu$m \citep{hall1978,scoville1983,beuther2010}. Using a rotation diagram analysis similar to that presented above, \citet{scoville1983} determined that the 8~km~s$^{-1}$ component contained total CO column densities of $N({\rm ^{12}CO})=(7.45\pm0.3)\times10^{18}$~cm$^{-2}$ and $N({\rm ^{13}CO})=(7.75\pm0.4)\times10^{16}$~cm$^{-2}$ at a rotational temperature of $T=150\pm10$~K, and that the  $-$17~km~s$^{-1}$ component contained total CO column densities of $N({\rm ^{12}CO})=(5.13\pm0.3)\times10^{18}$~cm$^{-2}$ and $N({\rm ^{13}CO})=(5.39\pm0.4)\times10^{16}$~cm$^{-2}$ at a rotational temperature of $T=150\pm30$~K. The spectral resolution of their observations ($\sim$7~km~s$^{-1}$) resulted in blending of the weaker $-$3~km~s$^{-1}$ component with the other two components, and precluded an independent analysis of this feature. \citet{scoville1983} also report components showing weak $^{12}$CO absorption at 30~km~s$^{-1}$ and $^{12}$CO $v=1$--0 emission at 20~km~s$^{-1}$, neither of which we detect in H$_2$O.

More recent 4.7~$\mu$m CO observations of BN were made using CRIRES \citep{kaufl2004} on UT1 at the VLT at $\sim$3~km~s$^{-1}$ spectral resolution \citep{beuther2010}. With telluric CO absorption affecting both the $-$3~km~s$^{-1}$ and 8~km~s$^{-1}$ components, these authors only analyzed the $-$17~km~s$^{-1}$ component, finding $N({\rm ^{13}CO})\approx5.7\times10^{16}$~cm$^{-2}$ at a rotational temperature of $T=112\pm20$~K, in agreement with the previous findings. However, \citet{beuther2010} did not detect CO absorption at 30~km~s$^{-1}$, and suggest that either the original detection must be a transient feature, or it was insignificant with respect to the continuum S/N level. Emission from both the $^{12}$CO and $^{13}$CO $v=1$--0 bands is reported at 20~km~s$^{-1}$, again in agreement with \citet{scoville1983}. While \citet{beuther2010} interpreted $^{12}$CO $v=1$--0 $R$(0) emission spatially extended along the slit as arising in an $r\sim1650$~AU circumstellar disk, millimeter continuum images show no evidence for such an extended structure in dust \citep{rodriguez2009,galvan-madrid2012,plambeck2013}, and it is unexpected that such a large structure would be retained around BN following its recent dynamical ejection at $\sim30$~km~s$^{-1}$.

Water absorption from the $\nu_3$ ro-vibrational band near 2.7~$\mu$m has also been observed toward BN before using the NASA Kuiper Airborne Observatory. \citet{knacke1991} detected absorption from both the $-$17~km~s$^{-1}$ and 8~km~s$^{-1}$ components in a handful of transitions out of low-lying states---albeit at rather low S/N---and estimated $N({\rm H_{2}O})=(2\pm1)\times10^{17}$~cm$^{-2}$, under the assumption that $T=150$~K. They also note the effects that a lower temperature would have, and give another estimate of $N({\rm H_{2}O})=(7.7\pm3.9)\times10^{16}$~cm$^{-2}$ if $T=70$~K. Our new results both confirm these findings and demonstrate the significant improvement achieved with the next generation of airborne observatories. 

Ideally, we would compare H$_2$O and CO abundances in all components that have been observed. However, the lack of a CO analysis for the 0.5~km~s$^{-1}$ component, the lack of H$_2$O data for the $-$17~km~s$^{-1}$ component, and the lack of H$_2$O emission at 20~km~s$^{-1}$, limit our options here. We can place a lower limit of $N({\rm H_{2}O})/N({\rm CO})\gtrsim 5\times10^{-4}$ in the $-$17~km~s$^{-1}$ component using only the column density measured in the 0$_{0,0}$ ground state of water. The 8~km~s$^{-1}$ component is the only one where both full CO and H$_2$O analyses exist, and there we find $N({\rm H_{2}O})/N({\rm CO})=5\times10^{-3}$. Both of these results are consistent with predictions of chemical models for cool, dense gas \citep[e.g., ][]{doty2002,hollenbach2009}, where most water has frozen out onto grains.

Previous low resolution ($R\sim670$) spectra of BN have in fact revealed broad absorption centered at 3.08~$\mu$m due to water ice \citep{knacke1982,smith1989}, and the narrow peak of this absorption feature requires that some amount of the ice be crystalline at $T>100$~K. The 3~$\mu$m ice absorption band presented in \citet{smith1989} has a peak optical depth of $\tau=1.78$ and a width of $\sim320$~cm$^{-1}$. Adopting a laboratory-measured integrated band strength for the O--H stretch mode in bulk H$_2$O ice of $2.0\times10^{-16}$~cm~per~molecule \citep{hagen1981}, this corresponds to an H$_2$O ice column density of $N({\rm H_2O_{ice}})\approx2.9\times10^{18}$~cm$^{-2}$. Assuming that the $-$17~km~s$^{-1}$ velocity component has similar excitation conditions and thus water abundance to the other two components, the ratio of water vapor to water ice toward BN is $N({\rm H_2O})/N({\rm H_{2}O_{ice}})\sim0.04$. This confirms the picture where water is predominantly in solid form.

BN has also been observed at $R\sim1400$ in the mid-IR with ISO-SWS \citep{gonzalez-alfonso1998,vandishoeck1998}. Both studies reported absorption and emission from the $\nu_2$ band of H$_2$O, but the large SWS aperture (14\arcsec$\times$20\arcsec\ at the shortest wavelengths) means that the region surrounding BN, including IRc2, was also contributing to the observed spectra. Our observations rule out BN as the source of the H$_2$O emission in those spectra, and rule out the 3 foreground components that we detect in H$_2$O absorption as giving rise to the H$_2$O absorption seen in those spectra.

\subsection{Nature of the foreground components}
The total visual extinction toward BN was estimated to be $A_V\approx17$~mag from observations of the 9.8~$\mu$m silicate absorption feature \citep{gezari1998}. Assuming $N_{\rm H}/A_V=1.8\times10^{21}$~cm$^{-2}$~mag$^{-1}$ \citep{bohlin1978}---where $N_{\rm H}\equiv N({\rm H}) + 2N({\rm H_2})$---this corresponds to a total hydrogen column density of $N_{\rm H}=3.1\times10^{22}$~cm$^{-2}$. The fractional abundance of water ice ($X({\rm H_{2}O_{ice}})\equiv N({\rm H_{2}O_{ice}})/N_{\rm H}$) averaged over all of the foreground gas is then $X({\rm H_{2}O_{ice}})\approx10^{-4}$, in agreement with predictions from photodissociation region (PDR) chemical models \citep[e.g.,][]{hollenbach2009}. It is not possible to determine which of the three foreground components contribute to the H$_2$O ice absorption signal, nor can any component be ruled out given the similar excitation conditions inferred from H$_2$O and CO. Again assuming that the $-$17~km~s$^{-1}$ component has similar $N({\rm H_2O})$ as the other two components, $X({\rm H_2O})\sim4\times10^{-6}$ in the foreground gas.

There is enough material along the line of sight that all three velocity components detected in H$_2$O absorption can potentially be associated with dense molecular gas \citep[$A_{V}\gtrsim5$~mag;][]{snow2006}. The depth into a cloud at which water ice becomes a major oxygen reservoir depends on the strength of the UV radiation field and gas density, but for much of the parameter space explored by \citet{hollenbach2009} $X({\rm H_{2}O_{ice}})$ exceeds $10^{-4}$ for $A_V\gtrsim3$~mag. In these same models $X({\rm H_{2}O})$ peaks at about 10$^{-7}$--10$^{-6}$ in a layer where water ice is being photodesorbed by the impinging far UV radiation \citep{melnick2011}. Additionally, $X({\rm CO})\sim10^{-4}$ at moderate cloud depths ($2\lesssim A_V\lesssim 4$~mag) before it freezes out onto grains at higher visual extinctions. The observed value of $X({\rm CO})$ averaged over all 3 cloud components is $\sim5\times10^{-3}$. The peak gas-phase H$_2$O abundance occurs farther into the cloud than the peak CO abundance, so it is expected that H$_2$O probes a region with lower gas temperature, in agreement with our findings.

The H$_2$O and CO observations suggest that the line of sight toward BN intersects three distinct clouds, all of which have fairly similar physical conditions ($n\sim10^4$--$10^5$~cm$^{-3}$, total $A_V\sim5$~mag, $T\sim60$--150~K). Given the limited range of $A_V$ available per cloud, the PDR models in \citet{hollenbach2009} favor a UV radiation field of order 100 times the average local interstellar radiation field \citep[parameterized as $G_0=1$;][]{habing1968} in order to roughly match the observed gas phase CO and H$_2$O ice abundances. These estimates are well below the extreme conditions ($G_0\gtrsim10^4$, $T\gtrsim500$~K) inferred from atomic and molecular emission line observations in the BN/KL and Trapezium regions \citep[e.g.,][]{melnick2011,goicoechea2015,goicoechea2015CII}, and our H$_2$O observations must be tracing a different physical component. Water emission observed toward the nearby Orion~KL region at high spectral resolution with {\it Herschel}/HIFI does not show any components that correspond with our observations \citep{melnick2010}. While the ``extended warm gas'' component matches the kinematics of our 8~km~s$^{-1}$ component, it has an H$_2$O column density over 100 times larger than what we have inferred. 
The H$_2$O and CO absorption could be associated with outflows from the region, including individual narrow streamers in the explosive outflow observed in CO emission \citep{bally2017}. Indeed, the lack of redshifted (with respect to the Orion molecular cloud) absorption toward BN suggests that it lies at a similar distance as the center of the explosive outflow, as would be expected if it played some part in triggering the event. Verification of this potential association would require a kinematic analysis of the CO streamers directly in front of BN. 
It is not trivial to relate the three foreground clouds seen toward BN with molecular emission previously observed in the region, and they may be associated with gas that is not easily traceable via emission in a crowded environment where many sources experience more extreme conditions.

\subsection{Lack of water in the vicinity of BN}

The radial velocity of BN has been best constrained by observations of hydrogen recombination lines arising from a hypercompact H~\textsc{ii} region \citep{rodriguez2009,plambeck2013}. In particular, \citet{plambeck2013} report $v_{\rm LSR}=23.2\pm0.5$~km~s$^{-1}$ from observations of the H30$\alpha$ line at 232~GHz. The CO $v=1$--0 emission detected by \citet{scoville1983} and \citet{beuther2010} is consistent with this velocity, hence both studies associate the emission with the immediate vicinity of BN. We detect neither emission nor absorption from H$_2$O at the BN radial velocity. The lack of H$_2$O absorption---and of CO absorption---at 23~km~s$^{-1}$, and the relatively low temperatures ($T\leq150$~K) inferred from H$_2$O and CO absorption at other velocities, indicates that there is no molecular foreground gas in close proximity to BN. 

If BN has a circumstellar disk it must be very compact, as mm continuum observations show the object to be a point source at 0\farcs15 ($\sim60$~AU) angular resolution \citep{rodriguez2009,galvan-madrid2012,plambeck2013}. It must also be oriented in such a way as to not cover the mid-IR continuum source, since that would presumably cause CO and H$_2$O absorption. This would seem to disfavor the disk interpretation of near-IR imaging polarimetry observations proposed by \citet{jiang2005}, and the large disk proposed by \citet{beuther2010}. However, the properties of the region giving rise to CO $v=1$--0 emission proposed by \citet{scoville1983} are consistent with the mm observations. It is possible that H$_2$O also resides in this region, but that its emission is too weak to be detected by our observations. Because quantitative speculation about such unobserved H$_2$O emission requires knowledge of several poorly constrained parameters (gas distribution, gas density, temperature, radiation field) we make no attempt to do so here. 


\section{Summary}
We have observed Orion BN at 6~$\mu$m and detected absorption from H$_2$O arising in cool ($T\sim70$~K), dense foreground clouds. The abundance ratios between water and CO, $N({\rm H_{2}O})/N({\rm CO})=5\times10^{-3}$, and between water vapor and water ice, $N({\rm H_2O})/N({\rm H_{2}O_{ice}})\sim0.04$, are indicative of a PDR environment where H$_2$O is released from ice mantles in a thin photodesorption layer. Neither H$_2$O absorption nor emission are detected at the systemic velocity of BN, adding further constraints to the size and orientation of any potential circumstellar disk.
The non-detection of H$_2$O in close proximity to BN adds new information regarding its current state, and should be accounted for in dynamical ejection models. For example, if BN was recently in a protostellar phase just before ejection, then either all of the dense gas must have been stripped away, or since accreted onto the star, or be oriented so as not to produce significant infrared absorption. In contrast, source I has retained a massive, Keplerian accretion disk \citep[e.g.,][]{hirota2014,ginsburg2018}. Finally, we note that BN may serve as a good sight line for studying molecular absorption at infrared wavelengths arising from cool, dense gas within the Orion molecular cloud. Over time, it may also enable the study of cloud structure in the plane of the sky as BN's high proper motion changes the foreground gas observed along the line of sight at different epochs.

Based on observations made with the NASA/DLR Stratospheric Observatory for Infrared Astronomy (SOFIA). SOFIA is jointly operated by the Universities Space Research Association, Inc. (USRA), under NASA contract NNA17BF53C, and the Deutsches SOFIA Institut (DSI) under DLR contract 50 OK 0901 to the University of Stuttgart. [Financial support for this work was provided by NASA through award \#04-0120 issued by USRA.]

\bibliographystyle{aasjournal}
\bibliography{indy_master}

\begin{thebibliography}{}
\expandafter\ifx\csname natexlab\endcsname\relax\def\natexlab#1{#1}\fi

\bibitem[{{Allen} \& {Burton}(1993)}]{allen1993}
{Allen}, D.~A., \& {Burton}, M.~G. 1993, \nat, 363, 54

\bibitem[{{Bally} {et~al.}(2017){Bally}, {Ginsburg}, {Arce}, {Eisner},
  {Youngblood}, {Zapata}, \& {Zinnecker}}]{bally2017}
{Bally}, J., {Ginsburg}, A., {Arce}, H., {et~al.} 2017, \apj, 837, 60

\bibitem[{{Bally} \& {Zinnecker}(2005)}]{bally2005}
{Bally}, J., \& {Zinnecker}, H. 2005, \aj, 129, 2281

\bibitem[{{Becklin} \& {Neugebauer}(1967)}]{becklin1967}
{Becklin}, E.~E., \& {Neugebauer}, G. 1967, \apj, 147, 799

\bibitem[{{Beuther} {et~al.}(2010){Beuther}, {Linz}, {Bik}, {Goto}, \&
  {Henning}}]{beuther2010}
{Beuther}, H., {Linz}, H., {Bik}, A., {Goto}, M., \& {Henning}, T. 2010, \aap,
  512, A29

\bibitem[{{Bohlin} {et~al.}(1978){Bohlin}, {Savage}, \& {Drake}}]{bohlin1978}
{Bohlin}, R.~C., {Savage}, B.~D., \& {Drake}, J.~F. 1978, \apj, 224, 132

\bibitem[{{Boonman} \& {van Dishoeck}(2003)}]{boonman2003}
{Boonman}, A.~M.~S., \& {van Dishoeck}, E.~F. 2003, \aap, 403, 1003

\bibitem[{{Chatterjee} \& {Tan}(2012)}]{chatterjee2012}
{Chatterjee}, S., \& {Tan}, J.~C. 2012, \apj, 754, 152

\bibitem[{{Clarke} {et~al.}(2015){Clarke}, {Vacca}, \& {Shuping}}]{clarke2014}
{Clarke}, M., {Vacca}, W.~D., \& {Shuping}, R.~Y. 2015, in Astronomical Data
  Analysis Software and Systems XXIV, ed. A.~R. {Taylor} \& J.~M. {Stil},
  Astronomical Society of the Pacific Conference Series

\bibitem[{{Cushing} {et~al.}(2004){Cushing}, {Vacca}, \&
  {Rayner}}]{cushing2004}
{Cushing}, M.~C., {Vacca}, W.~D., \& {Rayner}, J.~T. 2004, \pasp, 116, 362

\bibitem[{{Doty} {et~al.}(2002){Doty}, {van Dishoeck}, {van der Tak}, \&
  {Boonman}}]{doty2002}
{Doty}, S.~D., {van Dishoeck}, E.~F., {van der Tak}, F.~F.~S., \& {Boonman},
  A.~M.~S. 2002, \aap, 389, 446

\bibitem[{{Farias} \& {Tan}(2018)}]{farias2018}
{Farias}, J.~P., \& {Tan}, J.~C. 2018, \aap, 612, L7

\bibitem[{{Galv{\'a}n-Madrid} {et~al.}(2012){Galv{\'a}n-Madrid}, {Goddi}, \&
  {Rodr{\'{\i}}guez}}]{galvan-madrid2012}
{Galv{\'a}n-Madrid}, R., {Goddi}, C., \& {Rodr{\'{\i}}guez}, L.~F. 2012, \aap,
  547, L3

\bibitem[{{Gezari} {et~al.}(1998){Gezari}, {Backman}, \& {Werner}}]{gezari1998}
{Gezari}, D.~Y., {Backman}, D.~E., \& {Werner}, M.~W. 1998, \apj, 509, 283

\bibitem[{{Ginsburg} {et~al.}(2018){Ginsburg}, {Bally}, {Goddi}, {Plambeck}, \&
  {Wright}}]{ginsburg2018}
{Ginsburg}, A., {Bally}, J., {Goddi}, C., {Plambeck}, R., \& {Wright}, M. 2018,
  \apj, 860, 119

\bibitem[{{Goicoechea} {et~al.}(2015{\natexlab{a}}){Goicoechea},
  {Chavarr{\'{\i}}a}, {Cernicharo}, {Neufeld}, {Vavrek}, {Bergin}, {Cuadrado},
  {Encrenaz}, {Etxaluze}, {Melnick}, \& {Polehampton}}]{goicoechea2015}
{Goicoechea}, J.~R., {Chavarr{\'{\i}}a}, L., {Cernicharo}, J., {et~al.}
  2015{\natexlab{a}}, \apj, 799, 102

\bibitem[{{Goicoechea} {et~al.}(2015{\natexlab{b}}){Goicoechea}, {Teyssier},
  {Etxaluze}, {Goldsmith}, {Ossenkopf}, {Gerin}, {Bergin}, {Black},
  {Cernicharo}, {Cuadrado}, {Encrenaz}, {Falgarone}, {Fuente}, {Hacar}, {Lis},
  {Marcelino}, {Melnick}, {M{\"u}ller}, {Persson}, {Pety}, {R{\"o}llig},
  {Schilke}, {Simon}, {Snell}, \& {Stutzki}}]{goicoechea2015CII}
{Goicoechea}, J.~R., {Teyssier}, D., {Etxaluze}, M., {et~al.}
  2015{\natexlab{b}}, \apj, 812, 75

\bibitem[{{Gonzalez-Alfonso} {et~al.}(1998){Gonzalez-Alfonso}, {Cernicharo},
  {van Dishoeck}, {Wright}, \& {Heras}}]{gonzalez-alfonso1998}
{Gonzalez-Alfonso}, E., {Cernicharo}, J., {van Dishoeck}, E.~F., {Wright},
  C.~M., \& {Heras}, A. 1998, \apjl, 502, L169

\bibitem[{{Habing}(1968)}]{habing1968}
{Habing}, H.~J. 1968, \bain, 19, 421

\bibitem[{{Hagen} {et~al.}(1981){Hagen}, {Tielens}, \& {Greenberg}}]{hagen1981}
{Hagen}, W., {Tielens}, A.~G.~G.~M., \& {Greenberg}, J.~M. 1981, Chemical
  Physics, 56, 367

\bibitem[{{Hall} {et~al.}(1978){Hall}, {Kleinmann}, {Ridgway}, \&
  {Gillett}}]{hall1978}
{Hall}, D.~N.~B., {Kleinmann}, S.~G., {Ridgway}, S.~T., \& {Gillett}, F.~C.
  1978, \apjl, 223, L47

\bibitem[{{Hirota} {et~al.}(2014){Hirota}, {Kim}, {Kurono}, \&
  {Honma}}]{hirota2014}
{Hirota}, T., {Kim}, M.~K., {Kurono}, Y., \& {Honma}, M. 2014, \apjl, 782, L28

\bibitem[{{Hollenbach} {et~al.}(2009){Hollenbach}, {Kaufman}, {Bergin}, \&
  {Melnick}}]{hollenbach2009}
{Hollenbach}, D., {Kaufman}, M.~J., {Bergin}, E.~A., \& {Melnick}, G.~J. 2009,
  \apj, 690, 1497

\bibitem[{{Indriolo} {et~al.}(2015){Indriolo}, {Neufeld}, {DeWitt}, {Richter},
  {Boogert}, {Harper}, {Jaffe}, {Kulas}, {McKelvey}, {Ryde}, \&
  {Vacca}}]{indriolo2015exes}
{Indriolo}, N., {Neufeld}, D.~A., {DeWitt}, C.~N., {et~al.} 2015, \apjl, 802,
  L14

\bibitem[{{Jiang} {et~al.}(2005){Jiang}, {Tamura}, {Fukagawa}, {Hough},
  {Lucas}, {Suto}, {Ishii}, \& {Yang}}]{jiang2005}
{Jiang}, Z., {Tamura}, M., {Fukagawa}, M., {et~al.} 2005, \nat, 437, 112

\bibitem[{{K\"{a}ufl} {et~al.}(2004){K\"{a}ufl}, {Ballester}, {Biereichel},
  {Delabre}, {Donaldson}, {Dorn}, {Fedrigo}, {Finger}, {Fischer}, {Franza},
  {Gojak}, {Huster}, {Jung}, {Lizon}, {Mehrgan}, {Meyer}, {Moorwood}, {Pirard},
  {Paufique}, {Pozna}, {Siebenmorgen}, {Silber}, {Stegmeier}, \&
  {Wegerer}}]{kaufl2004}
{K\"{a}ufl}, H., {Ballester}, P., {Biereichel}, P., {et~al.} 2004, \procspie,
  5492, 1218

\bibitem[{{Knacke} \& {Larson}(1991)}]{knacke1991}
{Knacke}, R.~F., \& {Larson}, H.~P. 1991, \apj, 367, 162

\bibitem[{{Knacke} {et~al.}(1982){Knacke}, {McCorkle}, {Puetter}, {Erickson},
  \& {Kraetschmer}}]{knacke1982}
{Knacke}, R.~F., {McCorkle}, S., {Puetter}, R.~C., {Erickson}, E.~F., \&
  {Kraetschmer}, W. 1982, \apj, 260, 141

\bibitem[{{Kounkel} {et~al.}(2017){Kounkel}, {Hartmann}, {Loinard},
  {Ortiz-Le{\'o}n}, {Mioduszewski}, {Rodr{\'{\i}}guez}, {Dzib}, {Torres},
  {Pech}, {Galli}, {Rivera}, {Boden}, {Evans}, {Brice{\~n}o}, \&
  {Tobin}}]{kounkel2017}
{Kounkel}, M., {Hartmann}, L., {Loinard}, L., {et~al.} 2017, \apj, 834, 142

\bibitem[{{Luhman} {et~al.}(2017){Luhman}, {Robberto}, {Tan}, {Andersen},
  {Giulia Ubeira Gabellini}, {Manara}, {Platais}, \& {Ubeda}}]{luhman2017}
{Luhman}, K.~L., {Robberto}, M., {Tan}, J.~C., {et~al.} 2017, \apjl, 838, L3

\bibitem[{{McCall}(2001)}]{mccall2001}
{McCall}, B.~J. 2001, PhD thesis, The University of Chicago

\bibitem[{{Melnick} {et~al.}(2011){Melnick}, {Tolls}, {Snell}, {Bergin},
  {Hollenbach}, {Kaufman}, {Li}, \& {Neufeld}}]{melnick2011}
{Melnick}, G.~J., {Tolls}, V., {Snell}, R.~L., {et~al.} 2011, \apj, 727, 13

\bibitem[{{Melnick} {et~al.}(2010){Melnick}, {Tolls}, {Neufeld}, {Bergin},
  {Phillips}, {Wang}, {Crockett}, {Bell}, {Blake}, {Cabrit}, {Caux},
  {Ceccarelli}, {Cernicharo}, {Comito}, {Daniel}, {Dubernet}, {Emprechtinger},
  {Encrenaz}, {Falgarone}, {Gerin}, {Giesen}, {Goicoechea}, {Goldsmith},
  {Herbst}, {Joblin}, {Johnstone}, {Langer}, {Latter}, {Lis}, {Lord}, {Maret},
  {Martin}, {Menten}, {Morris}, {M{\"u}ller}, {Murphy}, {Ossenkopf}, {Pagani},
  {Pearson}, {P{\'e}rault}, {Plume}, {Qin}, {Salez}, {Schilke}, {Schlemmer},
  {Stutzki}, {Trappe}, {van der Tak}, {Vastel}, {Yorke}, {Yu}, \&
  {Zmuidzinas}}]{melnick2010}
{Melnick}, G.~J., {Tolls}, V., {Neufeld}, D.~A., {et~al.} 2010, \aap, 521, L27

\bibitem[{{Menten} {et~al.}(2007){Menten}, {Reid}, {Forbrich}, \&
  {Brunthaler}}]{menten2007}
{Menten}, K.~M., {Reid}, M.~J., {Forbrich}, J., \& {Brunthaler}, A. 2007, \aap,
  474, 515

\bibitem[{{Plambeck} {et~al.}(2013){Plambeck}, {Bolatto}, {Carpenter},
  {Eisner}, {Lamb}, {Leitch}, {Marrone}, {Muchovej}, {P{\'e}rez}, {Pound},
  {Teuben}, {Volgenau}, {Woody}, {Wright}, \& {Zauderer}}]{plambeck2013}
{Plambeck}, R.~L., {Bolatto}, A.~D., {Carpenter}, J.~M., {et~al.} 2013, \apj,
  765, 40

\bibitem[{{Richter} {et~al.}(2010){Richter}, {Ennico}, {McKelvey}, \&
  {Seifahrt}}]{richter2010}
{Richter}, M.~J., {Ennico}, K.~A., {McKelvey}, M.~E., \& {Seifahrt}, A. 2010,
  in Society of Photo-Optical Instrumentation Engineers (SPIE) Conference
  Series, Vol. 7735, Society of Photo-Optical Instrumentation Engineers (SPIE)
  Conference Series

\bibitem[{{Rodr{\'{\i}}guez} {et~al.}(2005){Rodr{\'{\i}}guez}, {Poveda},
  {Lizano}, \& {Allen}}]{rodriguez2005}
{Rodr{\'{\i}}guez}, L.~F., {Poveda}, A., {Lizano}, S., \& {Allen}, C. 2005,
  \apjl, 627, L65

\bibitem[{{Rodr{\'{\i}}guez} {et~al.}(2009){Rodr{\'{\i}}guez}, {Zapata}, \&
  {Ho}}]{rodriguez2009}
{Rodr{\'{\i}}guez}, L.~F., {Zapata}, L.~A., \& {Ho}, P.~T.~P. 2009, \apj, 692,
  162

\bibitem[{{Scoville} {et~al.}(1983){Scoville}, {Kleinmann}, {Hall}, \&
  {Ridgway}}]{scoville1983}
{Scoville}, N., {Kleinmann}, S.~G., {Hall}, D.~N.~B., \& {Ridgway}, S.~T. 1983,
  \apj, 275, 201

\bibitem[{{Smith} {et~al.}(1989){Smith}, {Sellgren}, \& {Tokunaga}}]{smith1989}
{Smith}, R.~G., {Sellgren}, K., \& {Tokunaga}, A.~T. 1989, \apj, 344, 413

\bibitem[{{Snow} \& {McCall}(2006)}]{snow2006}
{Snow}, T.~P., \& {McCall}, B.~J. 2006, ARA\&A, 44, 367

\bibitem[{{Tan}(2004)}]{tan2004}
{Tan}, J.~C. 2004, \apjl, 607, L47

\bibitem[{{van Dishoeck} {et~al.}(1998){van Dishoeck}, {Wright}, {Cernicharo},
  {Gonz{\'a}lez-Alfonso}, {de Graauw}, {Helmich}, \&
  {Vandenbussche}}]{vandishoeck1998}
{van Dishoeck}, E.~F., {Wright}, C.~M., {Cernicharo}, J., {et~al.} 1998, \apjl,
  502, L173

\bibitem[{{Young} {et~al.}(2012){Young}, {Becklin}, {Marcum}, {Roellig}, {De
  Buizer}, {Herter}, {G{\"u}sten}, {Dunham}, {Temi}, {Andersson}, {Backman},
  {Burgdorf}, {Caroff}, {Casey}, {Davidson}, {Erickson}, {Gehrz}, {Harper},
  {Harvey}, {Helton}, {Horner}, {Howard}, {Klein}, {Krabbe}, {McLean}, {Meyer},
  {Miles}, {Morris}, {Reach}, {Rho}, {Richter}, {Roeser}, {Sandell}, {Sankrit},
  {Savage}, {Smith}, {Shuping}, {Vacca}, {Vaillancourt}, {Wolf}, \&
  {Zinnecker}}]{young2012}
{Young}, E.~T., {Becklin}, E.~E., {Marcum}, P.~M., {et~al.} 2012, \apjl, 749,
  L17

\end{thebibliography}


\clearpage
\begin{deluxetable}{cccccccc}
\tabletypesize{\scriptsize}
\tablecaption{Absorption Line Parameters \label{tbl_measurements}}
\tablehead{\colhead{Transition} & \colhead{Wavelength} & \colhead{$E_{l}/k_{b}$} & \colhead{$v_{\rm LSR}$} & \colhead{FWHM} & \colhead{$\tau_{0}$} & \colhead{$\int\tau dv$} & \colhead{$N_l$}  \\
 & \colhead{($\mu$m)} & \colhead{(K)} & \colhead{(km~s$^{-1}$)} & \colhead{(km~s$^{-1}$)} &  & \colhead{(km~s$^{-1}$)} & \colhead{(10$^{15}$~cm$^{-2}$)} 
}
\startdata
1$_{1,1}$--0$_{0,0}$ & 6.116331 & 0 & 7.5 & 5.3$\pm$0.2 & 0.84$\pm$0.02 & 4.78$\pm$0.19 & 2.35$\pm$0.09 \\ 
3$_{1,2}$--3$_{0,3}$ & 6.113771 & 196.8 & 8.2 & 4.3$\pm$0.3 & 0.21$\pm$0.01 & 0.95$\pm$0.10 & 1.67$\pm$0.17 \\ 
3$_{2,1}$--3$_{1,2}$ & 6.075447 & 249.4 & 7.8 & 4.8$\pm$0.4 & 0.06$\pm$0.01 & 0.31$\pm$0.05 & 0.51$\pm$0.08 \\ 
\hline
1$_{1,1}$--0$_{0,0}$ & 6.116331 & 0 & 0.4 & 13.3$\pm$0.8 & 0.27$\pm$0.01 & 3.76$\pm$0.25 & 1.85$\pm$0.12 \\ 
3$_{1,2}$--3$_{0,3}$ & 6.113771 & 196.8 & 0.7 & 14.7$\pm$1.1 & 0.13$\pm$0.01 & 2.17$\pm$0.19 & 3.82$\pm$0.33 \\ 
3$_{2,1}$--3$_{1,2}$ & 6.075447 & 249.4 & 0.6 & 14.0$\pm$1.3 & 0.07$\pm$0.01 & 1.00$\pm$0.11 & 1.67$\pm$0.20 \\ 
\hline
1$_{1,1}$--0$_{0,0}$ & 6.116331 & 0 & $-$17 & 8.0$\pm$0.2 & 0.63$\pm$0.01 & 5.42$\pm$0.16 & 2.66$\pm$0.08 \\ 
\enddata
\tablecomments{Transition properties, parameters for the gaussian fits shown in Figure \ref{fig_spectra_fits}, and derived column densities.}
\end{deluxetable}
\normalsize


\clearpage

\end{document}